\documentclass[twocolumn]{aastex631}

\usepackage{float,graphicx,amsmath,multirow}
\usepackage{mhchem}
\usepackage{booktabs}
\usepackage{enumitem}
\usepackage{wrapfig}
\usepackage{threeparttable}
\usepackage{comment}

\begin{document}

\title{Exploring the Limits of Spectral Line Stacking in Spectral Line Data and Application Toward the Detection of Bulk \ce{^13C} Enrichment of Aromatics in TMC-1}


\author[0009-0007-2048-2907]{Miya Duffy}
\affiliation{Department of Chemistry, Massachusetts Institute of Technology, Cambridge, MA 02139, USA}

\author[0000-0002-8932-1219]{Ryan A. Loomis}
\affiliation{National Radio Astronomy Observatory, Charlottesville, VA 22903, USA}


\author[0009-0008-1171-278X]{Martin S. Holdren}
\affiliation{Department of Chemistry, Massachusetts Institute of Technology, Cambridge, MA 02139, USA}

\author[0000-0002-6667-7773]{Andrew Lipnicky}
\affiliation{National Radio Astronomy Observatory \\
520 Edgemont Road \\
Charlottesville, VA 22903, USA}

\author[0009-0005-1773-8460]{D. Archie Stewart}
\affiliation{Department of Chemistry, Massachusetts Institute of Technology, Cambridge, MA 02139, USA}

\author[0000-0002-0332-2641]{Gabi Wenzel}
\affiliation{Department of Chemistry, Massachusetts Institute of Technology, Cambridge, MA 02139, USA}
\affiliation{Center for Astrophysics \textbar{} Harvard \& Smithsonian, Cambridge, MA 02138, USA}

\author[0000-0003-2760-2119]{Ci Xue}
\affiliation{National Radio Astronomy Observatory, Charlottesville, VA 22903, USA}
\affiliation{NSF–Simons AI Institute for Cosmic Origins, Austin, TX, 78712, USA}

\author[0000-0002-9995-2700]{Annapoorani Hariharan}
\affiliation{Department of Chemistry, Massachusetts Institute of Technology, Cambridge, MA 02139, USA}

\author[0000-0002-0850-7426]{Ilsa R. Cooke}
\affiliation{Department of Chemistry, University of British Columbia, 2036 Main Mall, Vancouver, BC V6T 1Z1, Canada}

\author[0000-0001-9479-9287]{Anthony J. Remijan}
\affiliation{National Radio Astronomy Observatory, Charlottesville, VA 22903, USA}

\author[0000-0001-9142-0008]{Michael C. McCarthy}
\affiliation{Center for Astrophysics \textbar{} Harvard \& Smithsonian, Cambridge, MA 02138, USA}

\author[0000-0003-1254-4817]{Brett A. McGuire}
\affiliation{Department of Chemistry, Massachusetts Institute of Technology, Cambridge, MA 02139, USA}
\affiliation{National Radio Astronomy Observatory, Charlottesville, VA 22903}

\correspondingauthor{Miya Duffy, Brett A. McGuire}
\email{mfduffy@mit.edu, brettmc@mit.edu}

\begin{abstract}

The formation history of polycyclic aromatic hydrocarbons (PAHs) in the interstellar medium remains a topic of active debate, with proposed mechanisms ranging from high-temperature stellar ejecta processes to low-temperature chemistry within molecular clouds. Recently, the identification of small PAHs in the cold, dark cloud TMC-1 has provided some circumstantial evidence of the latter. If these are formed in situ, their isotopic ratios, particularly \ce{^12C}/\ce{^13C}, should reflect the local, bulk material of the molecular cloud; in contrast, PAHs formed in the circumstellar envelopes of evolved stars may show enhanced \ce{^13C} abundances. The expected radio signals of these \ce{^13C} substituted species will be faint, and thus we conducted a detailed proof-of-concept analysis examining whether spectral line stacking and matched filtering techniques can robustly retrieve signal from multiple singly substituted \ce{^13C} isotopologues in aggregate. We find that while retrieved signal decreases in the limit of spectral line confusion, false-positive detections are exceedingly improbable at the adopted 5$\sigma$ matched filter response detection threshold. We then demonstrate the technique on actual observational data of isotopologues of \ce{HC9N} toward TMC-1. Finally, we show using synthetic data that if laboratory rotational spectra of all singly substituted \ce{^13C} isotopologues of cyanonaphthalene and cyanopyrene isomers existed, current observations of TMC-1 would be sensitive enough to discriminate between a local, bulk \ce{^12C}/\ce{^13C} ratio and one with an enhanced \ce{^13C} abundance suggestive of inheritance.

\end{abstract}


\section{Introduction} \label{sec:intro}

Polycyclic aromatic hydrocarbons (PAHs) are widely dispersed across a variety of astrophysical environments \citep{peeters_astronomical_2011}. They are estimated to hold up to 10-25\% of the entire carbon budget of our Galaxy based on their assignment as molecular carriers of the 3.3, 6.2, 7.7, 8.6, and 11.2 micron spectral features known as the unidentified infrared bands (UIRs) \citep{tielens_interstellar_2008,allamandola_polycyclic_1985}. These spectral features are associated with the various vibrational modes of PAHs and are commonly used as observational evidence of the presence of PAHs in highly irradiated photo-dissociation regions \textcolor{black}{(PDRs)}, planetary nebulae, and young stellar objects \citep{joblin_pahs_2020}. The direct identification of individual interstellar PAHs (in most cases tagged with a cyanide (-CN) group), occurred much later through radio observations of their gas-phase rotational emission spectra toward the cold Taurus Molecular Cloud, TMC-1 \citep{burkhardt_discovery_2021,mcguire_detection_2021,wenzel_detection_2024,wenzel_detections_2025,wenzel_discovery_2025}. The detection of these individual species enables far more detailed investigations into their formation and evolution.  Because PAHs are such a large reservoir of reactive carbon, understanding these processes is key to understanding the cycle of carbon throughout star and planet formation.  Yet, even straightforward questions such as where and when these PAHs are formed remain.  One possibility for the CN-PAHs detected in TMC-1 is that they were inherited, formed by a previous generation of stars in circumstellar envelopes (CSEs), then ejected into the diffuse ISM and incorporated into TMC-1. Alternatively, the formation of PAHs may have occurred \textit{in-situ} in TMC-1, starting from the available materials already present in the cloud. It is also very likely that the truth of their formation consists of some combination of inheritance and \textit{in-situ} scenarios. 

\textcolor{black}{Isotopic composition can provide a powerful diagnostic tool to begin disentangling the origin environments of molecular species. \citet{steber_low_2025} recently measured and searched for singly substituted deuterium isotopomers of benzonitrile in TMC-1 in order to probe whether the isotopic content reflects the deuterium depletion that is expected for \textit{in situ} formation or enrichment, which would suggest possible inheritance from a previous UV-illuminated source. Their derived upper limit calculations suggest the D/H ratio of benzonitrile in TMC-1 \textcolor{black}{may} be consistent with that of other molecules residing there, \textcolor{black}{however, a straightforward link between the deuteration of PAHs in dark clouds and UV-illuminated regions has not yet been established, and it remains unclear whether this kind of deuterium enrichment holds true for PAHs in other interstellar regions.} In a similar vein,} the distribution of \ce{^13C} across the interstellar medium holds insights into the molecular processing history and reaction mechanisms responsible for forming a given molecule. This is the result of both how elemental carbon forms and various isotopic exchange reactions that are governed by zero-point energy differences. In the late stages of stellar evolution, stellar bodies lose mass through strong stellar outflows and winds. An estimated 90\% of the dust that is ejected into the interstellar medium (ISM) in our galaxy comes directly from these stellar mass loss processes \citep{karakas_dawes_2014}. In the production of \ce{^12C} and \ce{^13C}, the crucial nucleosynthetic period is the beginning phases of the asymptotic giant branch (AGB), where intense periods of dredge-up and mixing events alter the surface composition by incorporating internally processed material that has been partially exposed to hydrogen burning processes \citep{karakas_dawes_2014}. \ce{^13C} functions as a catalytic byproduct in the carbon-nitrogen-oxygen (CNO) cycle that occurs during stages of hydrogen burning and has been observed to be in higher quantities in the CSEs of a majority of stars that have been studied \citep{yan_direct_2023}. In an another study of 23 different circumstellar sources, \citet{milam_circumslar_2008} reported all observed carbon isotope ratios (\ce{^12C/^13C}) fell below the solar value of 89, with a majority exhibiting a ratio around 35, suggesting some level of \ce{^13C} enrichment in these regions compared to the solar value and that of the local ISM. This enrichment of \ce{^13C} in CSEs is a well-known phenomenon, and thus may help in the differentiation of molecular species of circumstellar versus interstellar origin. 

\textcolor{black}{The proposal for the circumstellar origin of PAHs is based on observations showing dust distribution in CSEs \citep{ziurys_chemistry_2006} and by analogy with combustion chemistry studies.} Terrestrial flame studies and astronomical observations suggest that astrophysically significant fractions of the bulk carbon atoms in carbon-rich CSEs are likely to be incorporated into molecules that form there, \textcolor{black}{some of which may be PAHs} \citep{feigelson_pah_1989}. \textcolor{black}{Although the similarities between combustion chemistry and circumstellar chemistry are encouraging, combustion science has not yet reached a consensus regarding chemical kinetic schemes for soot formation, so circumstellar PAH formation mechanisms are still largely unclear.} Successively larger PAHs may form via the consistent repetition of hydrogen abstraction followed by acetylene addition, i.e. the HACA mechanism \citep{joblin_pahs_2020}, or similar pathways. Additional computational studies have used gas-phase kinetic modeling for a set of reaction mechanisms previously used to study soot production in hydrocarbon flames. \citet{feigelson_pah_1989} modified and applied this model to circumstellar chemistry, concluding that a significant quantity of PAHs can in fact form in carbon-rich CSEs, given the density, initial acetylene fraction, and residence time in the 900-1100\,K temperature range are sufficiently high. For those detected in TMC-1, it may be the case that large carbonaceous grains are formed in CSEs, then subsequently seed the ISM with smaller PAHs as these larger structures are fragmented by interactions with interstellar radiation \citep{frenklach_formation_1989, allamandola_polycyclic_1985}. \textcolor{black}{With the continued uncertainties in what is known about soot formation, however, other studies have reported using similar models to \citet{feigelson_pah_1989} and predicted low PAH yields in AGB outflows \citep{cherchneff_polycyclic_1992} or even the preferential formation of aliphatic carbon species rather than aromatic \citep{martinez_prevalence_2019}.}

On the other hand, the most compelling evidence for \textit{in-situ} PAH formation in a cold molecular cloud environment continues to be the direct identification and quantification of the CN-PAHs cyanoindene, cyanonaphthalene, cyanoacenaphthylene, cyanopyrene, and cyanocoronene, along with the pure PAHs indene and phenalene in TMC-1 \citep{burkhardt_discovery_2021,mcguire_detection_2021,cernicharo_discovery_2024,wenzel_detections_2025,cabezas_discovery_2025}. Previous laboratory and theoretical studies have suggested PAHs of this (relatively) small size are unlikely to survive the harsh radiation conditions they would experience in their journey between circumstellar envelopes and cold molecular clouds like TMC-1 \citep{micelotta_polycyclic_2010,micelotta_polycyclic_2011}. Through a combination of laboratory and computational studies of the PAHs naphthalene, anthracene, and coronene with carbon atoms, \citet{krasnokutski_growth_2017} suggest that all small and large catacondensed PAHs react barrierlessly with atomic carbon and are likely to be efficiently destroyed by these interactions over a broad range of temperatures. However, more recent laboratory work has shown that survival mechanisms may indeed exist for certain derivatives of neutral PAHs. \citet{gatchell_survival_2021} showed that the reactive fragments of coronene resulting from high energy collisions with He atoms may survive at thermal equilibrium under ISM conditions. Recurrent fluorescence of the 1-cyanonaphthalene cation \citep{stockett_efficient_2023} and the indenyl cation \citep{bull_radiative_2025} has also been shown to \textcolor{black}{be an} effective and efficient stabilization mechanism under ISM conditions. 

Additional circumstantial evidence of interstellar PAH formation has recently been found in our Solar System based on samples returned from the primitive near-Earth carbonaceous asteroid Ryugu. Using 2-D gas chromatography with time-of-flight mass spectrometry, \citet{naraoka_soluble_2023} notably found fluoranthene and pyrene in the highest abundance in their samples from the 2019 Hayabusa2 touch-down operation on the asteroid Ryugu. The authors suggest that these species are likely to be presolar PAHs (i.e. formed in the interstellar medium), which were later incorporated into the parent body during accretion. A subsequent study conducted by \citet{sabbah_first_2024} analyzed the aromatic composition of another Ryugu sample, identifying aromatic species as large as 61 carbon atoms. Expanding on these works, \citet{zeichner_polycyclic_2023} utilized the sensitivity of terrestrial instruments to enable their investigation of the \ce{^13C} content of small PAHs from Ryugu. \textcolor{black}{Their carbon isotope clumping analysis revealed a small set of the sampled 3-ring PAHs like phenanthrene and anthracene, exhibited a \ce{^13C} isotopic content that was suggested to be indicative of high temperature circumstellar origins ($\sim$1000\,K). Conversely, 2- and 4-ring PAHs like naphthalene and pyrene, whose CN-derivatives have been observed in TMC-1 in high abundance, exhibited a carbon isotope content consistent with a cold formation environment ($\sim$10\,K), suggesting certain PAH structures may be linked to their interstellar formation environments.} 

Building on these findings, we now examine whether similar isotopic constraints can be applied to the PAHs detected in TMC-1. The solar system \ce{^12C}/\ce{^13C} ratio is estimated to be $\sim$\,89, which reflects the early carbon isotope ratio in the diffuse ISM. As a function of time, \ce{^13C} content is expected to increase, lowering the ratio in the local ISM which is now estimated to be $\sim$\,69 \citep{wilson_isotopes_1999}. Within a single source, additional variations due to differences in zero-point energies and self-shielding effects, among others, may also be at play. If these species form \textit{in situ}, their \ce{^13C} content should reflect the TMC-1 \ce{^12C/^13C} ratio. \textcolor{black}{A number of smaller and simpler carbon species and their \ce{^13C} isotopologues have been observed in TMC-1 and are thought to form \textit{in situ}. \citet{burkhardt_detection_2018} calculates isotopic ratios for each substitution position in \ce{HC5N} and \ce{HC7N}, reporting a weighted \ce{^12C/^13C} average value of 75 and 113, respectively. \ce{CCH} was investigated by \citet{sakai_abundance_2010}, finding that while the two carbon positions were not equivalent, the ratios were still quite large; [\ce{CCH}]/[\ce{C^13CH}] estimated to be larger than 170, and [\ce{CCH}]/[\ce{^13CCH}] exceeding 250. More recently, \citet{cabezas_detection_2025} reported an isotopic ratio of 93 for the hydrocarbon \ce{CH3C4H}. These results would suggest some level of \ce{^13C} dilution for these hydrocarbon species in TMC-1 compared to the local ISM, whereas significant \ce{^13C} enrichment would imply inheritance from CSE-processed material.}

The detection and quantification of \ce{^13C} substituted CN-PAHs, however, presents a substantial observational challenge.  While individual spectral lines of the known (CN)-PAHs in TMC-1 have been observed for the vast majority of these species, these lines are often at the limit of detectability even after hundreds of hours of integration \citep{wenzel_discovery_2025}.  Thus, many of these detections have been bolstered by the use of spectral line stacking and matched filtering to determine the aggregate emission of the molecule across multiple observed spectral lines \citep{loomis_investigation_2021}.  Here, we investigate the limits of these techniques to retrieve extremely weak signals, as well as whether they can be applied to multiple molecules simultaneously, thus retrieving, e.g., the aggregate emission from all \ce{^13C} substituted isotopologues of given (CN-)PAH.

\section{Methods}
\label{sec:methods}

A major bottleneck in the radio astronomical detection of additional PAHs in the ISM is a lack of the required laboratory rotational spectra. This becomes even more evident when it comes to singly-substituted \ce{^13C} isotopologues of these molecules. Since each hydrocarbon counterpart PAH possesses numerous unique carbon positions for substitution, each CN-substituted derivative results in a unique laboratory rotational spectrum.  They are often challenging to detect in natural abundance in the laboratory, or to synthesize through targeted enrichment. Thus, it is critical to ensure that there is a reasonable chance that they can be effectively used in conjunction with observations to enable either a detection or a meaningful limit, before undertaking substantial laboratory efforts.  

As mentioned earlier, spectral line stacking and matched filtering techniques have proven useful in enhancing the ability to robustly detect individual low-abundance species in observational data \citep{loomis_detecting_2018,loomis_investigation_2021}.  The use of these stacking and filtering techniques on aggregate molecular emission from multiple species in radio spectra, however, has not been demonstrated nor tested before. Indeed, to our knowledge, a thorough exploration of the limits of our ability to reliably extract signal (from individual species or in aggregate) using these techniques -- minimizing false positives -- as observational data become increasingly line rich has also not been explored. Thus, we first identify the limitations of these techniques as a function of both the noise level of the astronomical observations and the density of interfering spectral lines.

Then, we investigate an expansion of these techniques in an attempt to detect the aggregate signal from \emph{all} \ce{^13C} singly-substituted isotopomers of a given molecule. We demonstrate proof-of-concept using \ce{HC9N} observations, for which the necessary isotopic laboratory data are known \citep{mccarthy_experimental_2000}. We then use calculated spectra for the PAHs 1- and 2-cyanonaphthalene and 1-, 2-, and 4-cyanopyrene to conduct a preliminary analysis and explore whether our current observations would be sensitive enough to enable a detection or inform meaningful constraints.

\subsection{Spectral Line Stacking and Matched Filtering}
\label{subsec:stacking}

\begin{figure*}
    \centering
    \includegraphics[width=1\linewidth]{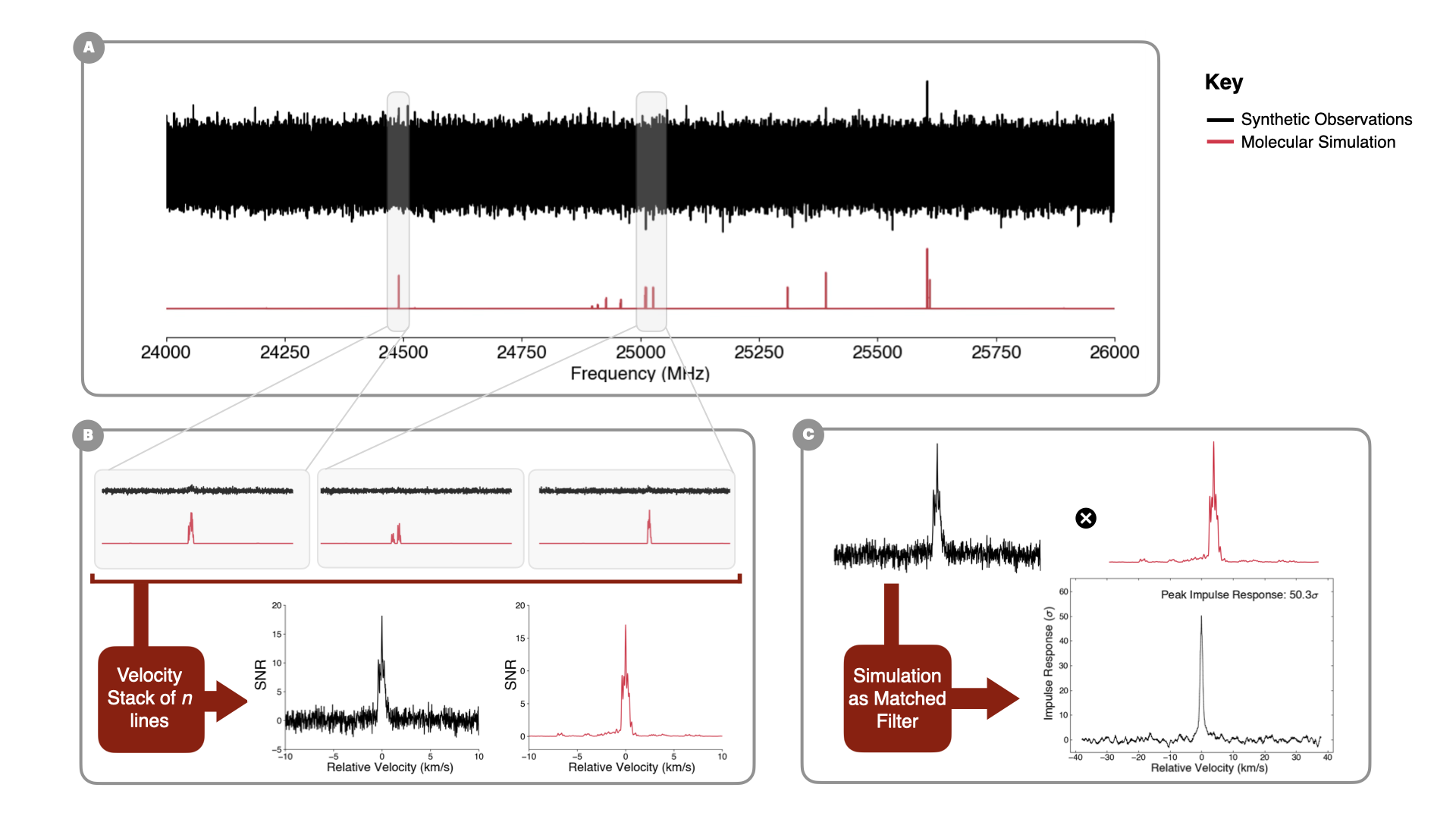}
    \caption{Schematic diagram of our spectral line stacking and matched filtering procedures. First, (a) shows relevant spectral regions of the \textcolor{black}{(synthetic)} observational data (black) and an initial spectral simulation (red). Each spectral window is then weighted based on the signal-to-noise \textcolor{black}{ratio} of each region to create a velocity stack of all relevant molecular rotational transitions (b). Finally, the stacked simulation is used as a matched filter and is cross-correlated with the stacked data to visualize the statistical significance of a given detection (c).}
    \label{fig:stacking}
\end{figure*}

Signal stacking is a technique used across a variety of research areas, such as seismology and oceanography \citep{borges_oliveira_cyclic_2021,dumas_new_2023}, as a method of aggregating multiple weaker signals in noisy data to ultimately recover a higher signal-to-noise ratio (SNR) detection. In astrochemistry, this technique has been applied to molecular rotational transitions, stacking all predicted lines from a given molecule to retrieve a single signal representing the total molecular signal (see, e.g., \citealt{langston_detection_2007,loomis_detecting_2018}). The ability to retrieve the total aggregate signal of a molecule provides the opportunity to look for less abundant species, expanding the pool of possible molecules that can be detected given the laboratory data is available. The relevant spectral regions \textcolor{black}{(i.e. all the rotational transitions that overlap with the frequency range of the observational dataset that will be included in the final stack)} used in a stack are determined by first simulating the rotational spectrum of a target species under the excitation conditions of the astronomical source. These regions are then extracted and re-gridded into velocity space with the center of each signal peak aligned to 0\,km\,s$^{-1}$. A SNR-weighted average is then used to produce a stacked spectrum in velocity space with substantially higher SNR than the individual lines, representing the total rotational signature of that species in the data. The same weights are then applied to identical chunks of the simulated spectrum of a molecule, and a stack is performed on those data as well, ultimately yielding a simulation of the predicted stacked velocity spectrum. The goal is then to determine the statistical evidence that this hypothesized signal is in agreement with the observations. \textcolor{black}{A schematic diagram of this procedure can be seen in Fig. \ref{fig:stacking}}. 

To do so, the stack of the simulated molecular rotational spectrum is used as a matched filter \citep{north_analysis_1963}. In the filtering process, the predicted stack of a simulated molecule is cross correlated with the observational stack, yielding an impulse response. Normalizing this impulse response such that the RMS noise level is set to 1 yields a result in units of $\sigma$, meaning that the impulse response represents the statistical confidence level that the observations and simulation match. If the molecular signal is absent in the observations or is too weak to be recovered, both the stack and the filter result will simply average to noise. Matched filters have been shown to provide an additional boost in SNR when used in conjunction with stacking on astronomical spectroscopic data, optimally extracting the statistical significance of a detection when the shape of the target signal is known; the application of a matched filter increased SNR by up to 50-60\% in both synthetic and real interferometric data test cases \citep{loomis_detecting_2018}. The GOTHAM (Green Bank Telescope (GBT) Observations of TMC-1: Hunting Aromatic Molecules) project has previously adopted a threshold for claiming a detection of 5$\sigma$ \citep{loomis_investigation_2021}. An impulse above this threshold indicates that the probability of this result being due solely to statistical variation is only 0.00003\% assuming Gaussian white noise and no competing signals. 
 
\subsection{GOTHAM Dataset}
\label{subsec:GOTHAM}

The GOTHAM dataset is a large broadband spectral line survey of TMC-1 using the 100\,m Robert C. Byrd Green Bank Telescope (GBT) with close to complete frequency coverage of 6--35\,GHz at a spectral resolution of 1.4\,kHz, \textcolor{black}{corresponding to a velocity resolution of 0.05-0.02\,km\,s$^{-1}$, depending on frequency}. As described in \citet{loomis_investigation_2021}, the spectral stacking and filtering method used in this work has been primarily demonstrated and tested on a relatively line sparse spectral dataset. Above 5$\sigma$, the line density in GOTHAM observations is approximately 0.05 lines per MHz (one line every 20\,MHz). The average full width at half-maximum (FWHM) line width of an emission line feature is \textcolor{black}{0.4\,km\,s$^{-1}$, which corresponds to 8\,kHz at 6\,GHz and 47\,kHz at 35\,GHz.} This results in, on average, one line every $\sim$500--3000 FWHM (i.e., extremely line-sparse). Even so, signal from hundreds of unique molecular species are present, from well-known molecules with bright emission signals \citep{mcguire_detection_2021} to those whose presence has been seen largely through the stacking of transitions which are mostly at or below the noise level \citep{sita_discovery_2022,remijan_astronomical_2023} i.e., not included in the prior line density calculations. This implies the potential presence of substantial numbers of additional, potentially interfering, transitions from as yet unidentified species below the current noise level. Indeed, \citet{mcguire_detection_2021} crudely estimated the line density in the GOTHAM dataset as a function of peak intensity, inferring a line density of 1 per 100 FWHM for all lines with an intensity $\geq$0.1\,mK (about 10-100$\times$ below the current GOTHAM noise level). Here, we wish to assess the potential for this ``hidden" line density to interfere with the recovery of our target signal.

While we ultimately wish to use our expanded methodologies of aggregate isotopologue signal recovery on the GOTHAM dataset, as an initial step, we benchmarked and stress tested them on a dataset with well constrained properties.  We first generated a set of synthetic observations very closely matching the spectral and noise properties of GOTHAM, but free from any molecular signal. Variations in local RMS noise levels across the full spectral range of GOTHAM were captured and recreated as white, Gaussian noise in the corresponding sections of the synthetic data. To assess signal recovery limits, we then injected simulated molecular spectral lines (\S\ref{subsec:methods_qc}) into this synthetic data, such that we then knew the ground-truth signals (both in terms of line density and their exact positions and intensities) present in the data. 

\subsection{Quantum Chemical Calculations and Rotational Spectra}
 \label{subsec:methods_qc}

For this analysis, we required calculated spectra for both of our target PAH species (all cyanonaphthalene and cyanopyrene isomers \textcolor{black}{and their singly substituted \ce{^13C} \textcolor{black}{isotopologues}}), as well as for an inventory of \textcolor{black}{arbitrarily generated }molecules which \textcolor{black}{are used to }represent the interfering transitions convoluting the (synthetic) observational spectrum.  

\subsubsection{PAH Spectra} The parent (\ce{^12C}) isotopologues of 1- and 2-cyanonaphthalene and 1-, 2-, and 4-cyanopyrene have been well-characterized in the laboratory \citep{mcnaughton_laboratory_2018,wenzel_detections_2025}, but their singly \ce{^13C} substituted isotopologues have not yet been studied. Therefore, rotational constants and permanent electric dipole moments for all possible single substitution positions were calculated using Gaussian 16 \citep{frisch_gaussian16_2016} at the M06-2X/6-31+G(d) level of theory, which has been shown to provide excellent estimates for a reasonable computational cost, especially on such large systems \citep{lee_bayesian_2020}. If experimentally determined rotational constants for the parent isotopologue are known, the accuracy of calculated rotational constants for isotopologues can be dramatically enhanced by applying a scaling factor determined from the ratio of the experimentally determined constants to those calculated for the parent at the same level of theory and basis set \citep{wenzel_laboratory_2025}. We derived and applied such scaling factors for this study, using experimentally determined rotational constants for the cyanonaphthalene \citep{mcnaughton_laboratory_2018} and cyanopyrene isomers \citep{wenzel_detection_2024,wenzel_detections_2025}. Centrifugal distortion constants and the \ce{^14N} nuclear electric quadrupole hyperfine coupling constants were assumed to be identical to the parent species. Ultimately, laboratory accuracy will still be required to enable a robust astronomical search, however, these constants are sufficiently accurate to ensure an almost identical spectral pattern robust enough to use in this proof-of-concept study. 
 
Synthetic rotational spectra were generated by simulating all species under the specific physical conditions of TMC-1 based on original detections of the unsubstituted parent molecules \citep{mcguire_detection_2021, wenzel_detection_2024,wenzel_detections_2025}, with the exception that a single velocity component, rather than the four nearly overlapping components seen in GOTHAM, was used for simplicity. Then, to facilitate a stacking and filtering analysis of the aggregate signal of all singly substituted isotopologues for a given isomer, a single spectrum was generated by summing \textcolor{black}{all isotopologues along a shared frequency axis.}\footnote{Isotopologues are molecular species that differ in their isotopic composition (i.e., number of isotopic substitutions), while isotopomers have the same number of each isotope but differ only in the specific substitution positions.} The result was a single aggregate spectrum per parent isomer. All aggregate isomer spectra for a single PAH molecule (3 and 2 for cyanopyrene and cyanonaphthalene, respectively) were then summed to yield a single rotational spectrum prediction for the total expected signal \textcolor{black}{from all possible singly substituted \ce{^13C} isotopomers at any given \ce{^12C}/\ce{^13C} ratio.} 

\subsubsection{Synthetic Interfering Lines} \label{subsec:interfering_lines} To manipulate the line density of the observational spectrum, a multitude of additional synthetic rotational spectra were generated to inject into the observational simulations. To generate the spectra representing a variety of asymmetric molecular geometries, principal rotational constants ($A$,$B$,$C$) were randomly generated with a uniform distribution over $\kappa$ values and $C/A$ ratios. For each randomly generated pair of $\kappa$ and $C/A$ values, a value for $B$ was randomly chosen from a uniform distribution from 1 GHz to 25 GHz (to ensure a reasonable number of features appeared in the range of the GOTHAM data), and then from these three values a unique set of $A$, $B$, and $C$ constants was derived. For each set of rotational constants, electric dipole components were randomly assigned boolean (i.e. 0 or 1) values. Each synthetic rotational spectrum was then generated based on these six randomly generated values: $A$, $B$, $C$, $\mu_a$, $\mu_b$, and $\mu_c$.

In total, 10,000 \textcolor{black}{catalogs} were generated up to 40\,GHz (sufficient to cover the GOTHAM data frequency range) using Herb Pickett's SPFIT/SPCAT software to calculate catalogs of rotational transitions \citep{pickett_fitting_1991}. We then used samplings of these catalogs to inject controlled, known spectral lines into the synthetic observations described above. Any integer number of total lines can be input into the program with corresponding intensity values in units relative to the local RMS for each frequency chunk, yielding a completely synthetic spectrum that incorporates real molecular signal while also being tunable to local RMS values. \textcolor{black}{The total number of lines is based on the desired line density of the outputted spectrum. With higher line densities, a greater number of the 10,000 synthetic catalogs are used.}This resulted in the production of synthetic spectra with arbitrary known line densities at arbitrary intensity thresholds. The use of catalogs generated using realistic sets of rotational constants ensures that characteristic molecular signal patterns (harmonic progressions, etc.) are also present in the data. The intensities of predicted rotational spectra were independently modulated by simulating species at different column densities corresponding to a given isotope ratio based on the total abundance of the parent.


\section{Results}
\label{sec:results}

\begin{figure*}[tbh!]
    \centering
    \includegraphics[width=1\linewidth]{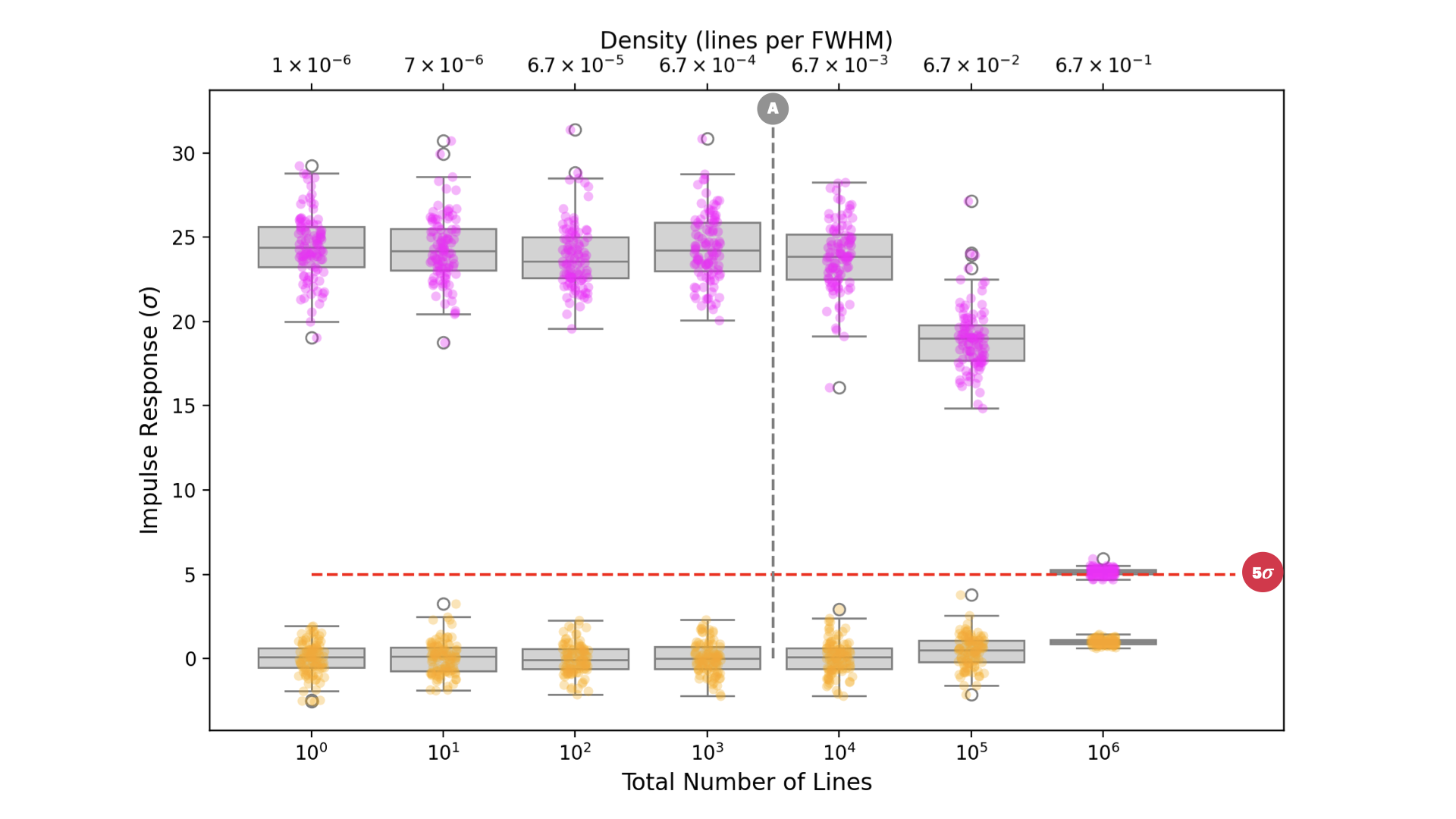}
    \caption{\textcolor{black}{Box and whisker plot of} line density impulse response distributions for false positive and false negative detection rates. \textcolor{black}{The gray box represents the inter-quartile range (IQR), the central line indicates the median, and the whiskers extend to the minimum and maximum values within 1.5×IQR. Open gray circles represent statistical outliers.} Each individual stack for the dataset presented in pink includes the simulated spectrum of unsubstituted 1-cyanonaphthalene at the observed column density of $N_T = 7.5\times10^{11}$\,cm$^{-2}$, \textcolor{black}{and molecular signal from synthetic catalogs at each given line density} showing false negative detection rates. Individual stacks in orange include only white Gaussian noise representative of the corresponding local RMS of GOTHAM observations \textcolor{black}{along with molecular signals at the line density shown along the x-axis}, showing the false positive detection rates as a function of line density. \textcolor{black}{The vertical dashed line (A) represents the line density above the noise level at 1$\sigma$ in GOTHAM observations.} The horizontal dashed red line is the adopted 5$\sigma$ impulse response threshold of detection.}
    \label{fig:false-rates}
\end{figure*}

We first performed tests on the synthetic observations to determine the rate of false positive and false negative detections using the velocity stack and matched filter method with respect to the intensity of the molecular signals and varying line densities. After showing the statistical validity in this way, we performed proof-of-concept analyses for the ability to retrieve aggregate \ce{^13C} signal in the GOTHAM data, using \ce{HC9N} as a test case. Finally, with assurances against false positives and the demonstrated ability to retrieve aggregate signal effectively, we then examined the ability of these methods to retrieve aggregate signal of cyanonaphthalene and cyanopyrene isotopomers in GOTHAM data under a range of putative \ce{^12C}/\ce{^13C} ratios, if laboratory data become available.

\subsection{False Positives/Negatives and Line Density}
\label{subsec:false}

To assess the likelihood of false-positive or false-negative detections from stacking and filtering as a function of line density, 100 synthetic spectra were generated as described in \S\ref{subsec:interfering_lines} for seven different line densities spanning from 10$^{-6}$ lines per full-width half-maximum (FWHM) up to 0.5 lines per FWHM (i.e., completely line-confused). The intensities of these lines were set to be equal to 1$\sigma$ of the local RMS value. In this way, they represent the ``worst-case" observational scenario, where there are as many lines of as high an intensity as possible lurking unseen beneath the noise to interfere. \textcolor{black}{An in-depth analysis of the spectral line density of the GOTHAM dataset is presented in Appendix \ref{appendix:linedensity}. }

\subsubsection{False Positives} To test for false positive rates, we attempted to retrieve signal from 1-cyanonaphthalene using our standard spectral stacking and matched filtering analysis in each of the 100 synthetic spectra at each of the seven line densities, assuming a single velocity component at $v_{lsr} = 5.8$\,km\,s$^{-1}$, $T_{ex} = 8.9$\,K, \textcolor{black}{$\Delta V = 0.4$\,km\,s$^{-1}$,} and a total column density of $N_T = 7.35\times10^{11}$\,cm$^{-2}$, matching the original detection of 1-cyanonaphthalene in TMC-1 by \citet{mcguire_detection_2021}. With over 1500 transitions within the range of the GOTHAM data, 1-cyanonaphthalene provided a large enough pool of total lines to stack in order to effectively investigate false positives. The results are shown in Fig.~\ref{fig:false-rates}. The orange dots in each box and whisker plot represent the retrieved impulse response from 1-cyanonaphthalene in each run, given that we know it is not present in the synthetic spectrum.  As expected, the retrieved values show a tight distribution centered around 0$\sigma$ for the vast majority of line densities. As shown in Fig.~\ref{fig:gauss_pos}, each of these distributions is well described by a Gaussian centered around 0, as expected for an attempt to retrieve a signal that is not present from data that is largely white Gaussian noise. As line-confusion is approached, the distribution shifts toward and collapses around 1$\sigma$. This is expected behavior, as the stack procedure will result in a narrow range of values in all velocity channels, which when normalized to have an RMS of 1 will also result in a retrieved impulse response of 1. This is shown in \textcolor{black}{Equations }\ref{eqn:snr1} and \ref{eqn:snr2}:  

\begin{equation}
    \text{SNR}_{\text{app}} = \frac{P_{\text{sig,meas}}}{P_{\text{noise,tot}}} = \frac{S + N_c}{N_w + N_c}
    \label{eqn:snr1}
\end{equation}
Where $S$ is the true signal, $N_c$ is the correlated noise (e.g., line forest contamination), and $N_w$ is the white noise. In the limit of high line density where $N_c \gg N_w$ and $S \to 0$, the $\text{SNR}_{\text{app}}$ converges to:
\begin{equation}
    \lim_{N_c \to \infty} \frac{S + N_c}{N_w + N_c} = \frac{N_c}{N_c} = 1
    \label{eqn:snr2}
\end{equation}
From this analysis, we conclude that the retrieved significance value from stacking analyses cannot be artificially inflated to provide a false-positive detection due to hidden spectral lines beneath the noise.

\subsubsection{False Negatives} Next, we investigated whether stacking and filtering becomes less useful as line density increases toward line confusion.  To do so, we injected signal from 1-cyanonaphthalene into each of 100 synthetic spectra at each of the seven line densities, again assuming a single velocity component at $v_{lsr} = 5.8$\,km\,s$^{-1}$, $T_{ex} = 8.9$\,K, \textcolor{black}{$\Delta V = 0.4$\,km\,s$^{-1}$,} and a total column density of $N_T = 7.35\times10^{11}$\,cm$^{-2}$, matching the original detection by \citet{mcguire_detection_2021}.  We then attempted to retrieve the signal from 1-cyanonaphthalene.  The results are shown as the pink dots in Fig.~\ref{fig:false-rates}. At low line densities, we achieve a robust $\sim$30$\sigma$ retrieval. This begins to fall off starting a line density $>$0.005 lines per FWHM, eventually falling below our 5$\sigma$ threshold at line confusion. This is expected behavior given \textcolor{black}{Equation }\ref{eqn:snr1} where as we approach a limit of large $N_c$ and there is true signal present in the data, there is a reduction in the apparent SNR of the recovered signal which can be seen in the right side of \textcolor{black}{Equation }\ref{fig:false-rates}. We note that the retrieved signal does not collapse to 1$\sigma$ as before, because there is still real signal in the spectrum at this point, however the contrast between real signal and the baseline -- the normalized ``noise" which is largely interloper signal -- becomes small.

\subsubsection{Specific Effects on GOTHAM Retrievals} The GOTHAM observations are estimated to have a line density of $\sim$0.005 lines per FWHM at 3$\sigma$. The vertical dashed line, A, in \textcolor{black}{Fig.}\ref{fig:false-rates} shows the point at which we'd expect to see a drop-off in significance of the recovered signal. Based on the results shown in Fig.~\ref{fig:false-rates}, we estimate that this density may be slightly depressing the overall significance of our retrieved impulse responses. That said, there are a number of caveats and cautions.  

First, as noted by \citet{mcguire_detection_2021} when they originally performed the analysis, the line density estimate below the noise level in GOTHAM was entirely empirical, using a power-law extrapolation from the density of lines above the noise. There is no compelling physical reason to believe such a functional form is valid, and thus the comparison point of 0.01 lines per FWHM \textcolor{black}{as the line density below the noise level in the observations} is highly speculative.  

Second, as also discussed in \citet{mcguire_detection_2021} and \citet{loomis_investigation_2021}, while the stacking and filtering results are dominated by actual molecular signal, there is a non-zero contribution of red noise that may become more impactful for extremely weak signals. The largest source of this red noise is believed to have been correlated noise on a few channel basis due to the regridding procedures used in early data reductions of GOTHAM. This has been addressed in the most recent reduction and first public release of the entire dataset, and so red noise contributions going forward are expected to be \textcolor{black}{negligible} \citep{xue_molecular_2025}.

Third, we note here that this analysis only addresses interference from buried interfering lines.  Interference from signals $>$1$\sigma$ remains a possibility.  This is addressed using standard signal rejection methods and manual inspection of the most highly weighted spectral chunks, as discussed in prior publications \citep{mcguire_detection_2021}. 

\paragraph{Takeaway Message} In light of the results of these analyses, we conclude that signals from our stacking and matched filtering analysis cannot be artificially inflated by a high density of buried spectral lines, but will instead be depressed by those interfering transitions at densities approaching line confusion. Any false-positives from stacking and filtering analyses, therefore, would most likely be caused by incomplete filtering of interfering signals above the noise level in the observations.

\subsection{\ce{^13C} Isotopologues of \ce{HC9N} in TMC-1}
 \label{subsec:hc9n}
To validate the methodology for retrieving an aggregate signal from multiple isotopologues, we applied the procedures outlined above to \ce{HC9N} using the GOTHAM data as a test case. The high-resolution rotational spectra of all nine \ce{^13C} isotopologues were reported in \citet{mccarthy_experimental_2000}. A detailed Markov-Chain Monte Carlo (MCMC) fitting and analysis of each individual species, along with astrochemical implications, will be presented in \citet{burkhardt_hc9n_2026}.  

\begin{figure}
    \centering
    \includegraphics[width=1\linewidth,trim=15cm 0cm 15cm 0cm, clip]{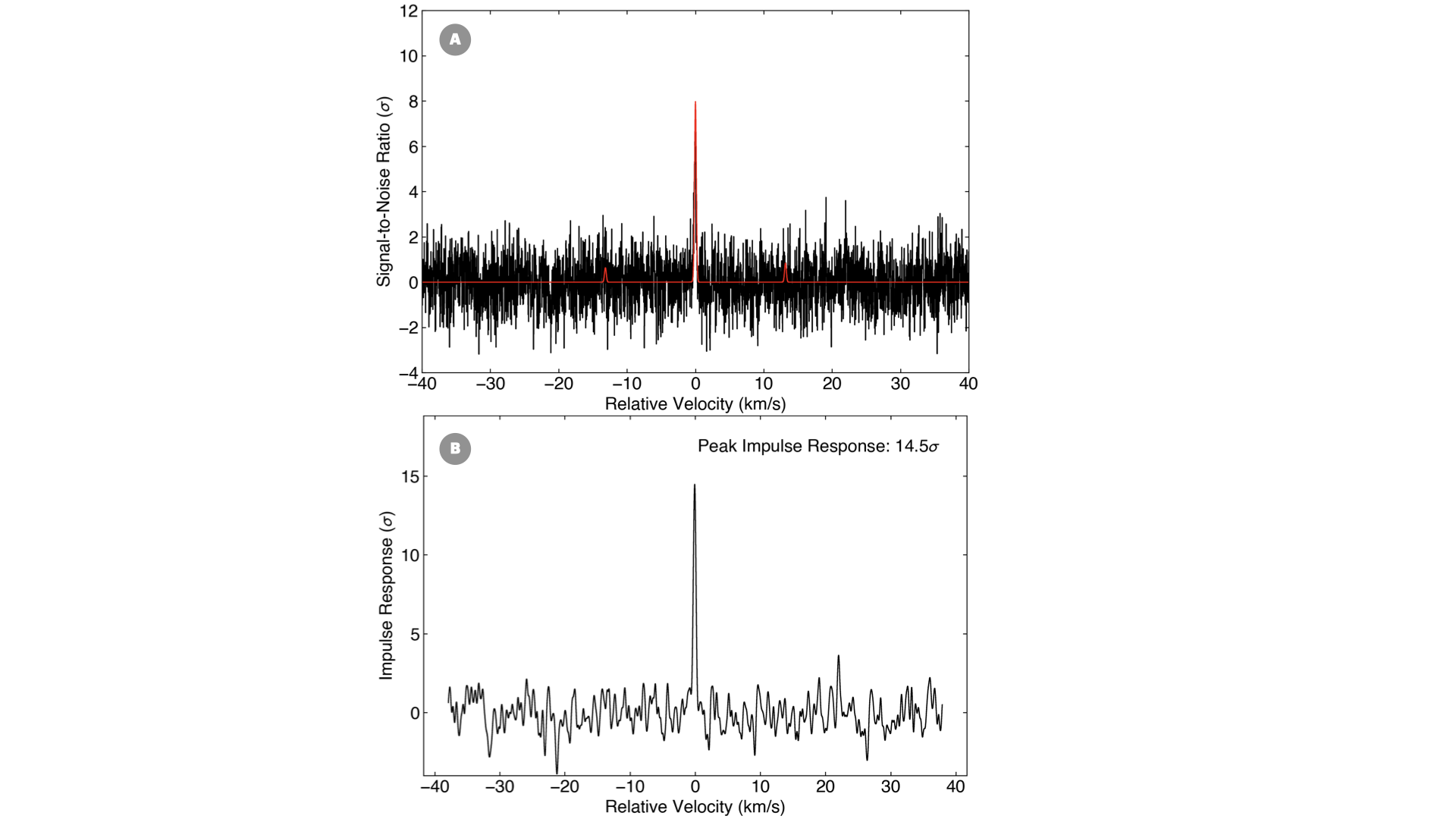}
    \caption{Velocity stack \textcolor{black}{(A)} and matched filter \textcolor{black}{(B)} spectra of \ce{HC9N} \textcolor{black}{\ce{^13C} isotopomers}. The intensity scales are the signal-to-noise ratios (SNR) of the response functions when centered at a given velocity. Velocity is relative to the systemic velocity of TMC-1 at 5.8 km\,s$^{-1}$. The stacked spectrum from the most recent GOTHAM data \textcolor{black}{is} shown in black. The stacked \ce{HC9N} \textcolor{black}{isotopomer} simulation is shown overlaid in red.}
    \label{fig:stack}
\end{figure}

Here, we assume uniform source parameters and column densities for all nine isotopologues, using the same single-component linewidth, velocity, and source size as above, and adopting  an excitation temperature of $8.7$\,K and a total column density of $N_T = 4.32\times10^{12}$\,cm$^{-2}$ from our fitting of \ce{HC9N} \citep{xue_molecular_2025}. Following the procedure outlined in \S\ref{subsec:methods_qc}, the spectra of all nine isotopologues were then simulated under these conditions and summed to produce an aggregate simulated \textcolor{black}{spectrum}. Each isotopomer simulation was based on a single velocity component model rather than 4 components for simplicity. With the loss of information in this simplification, the results may represent a lower limit for aggregate signal recovery. The aggregate spectrum of all nine isotopomers was then used to perform the spectral stacking and matched filtering analysis. Strong signal was seen in the spectral stack and matched filter (14.5$\sigma$; Fig.~\ref{fig:stack}). Because we did not perform detailed MCMC fitting to the individual species, we empirically scaled the bulk \ce{^12C}/\ce{^13C} ratio to a value of 130 to match the spectral stack intensity by eye. This ratio should be viewed as a loose estimate; a more rigorous analysis will be presented later in \citet{burkhardt_hc9n_2026}. For the purposes of this study, however, it is clear that retrieval of aggregate signal from a collection of isotopologues is feasible.

\subsection{Cyanonaphthalene and Cyanopyrene}
 \label{subsec:pahs}
With the ability to recover weak signals without generating false positives, we then investigated the lowest detectable column density of these PAHs while still retrieving a 5$\sigma$ impulse response. The laboratory spectra for the \ce{^13C} isotopologues of the cyanonaphthalenes and cyanopyrenes are not yet available. We have therefore conducted a study examining what ratios of \ce{^12C}/\ce{^13C} would result in a detection in the GOTHAM data, were the laboratory spectra available.  

For both PAH groups, we combine the procedures outlined above in \S\ref{subsec:false} and \S\ref{subsec:hc9n} by simulating the spectra of all \ce{^13C} isotopologues, using the constants calculated in \S\ref{subsec:methods_qc}, at a range of \ce{^12C}:\ce{^13C} ratios from 10--100.  We then injected these signals into synthetic observations and determined whether a robust ($>$5$\sigma$) aggregate signal was retrievable at each \ce{^12C}:\ce{^13C} ratio, repeating the experiment 100 times at each ratio.

\begin{figure}
    \centering
    \includegraphics[width=1\linewidth]{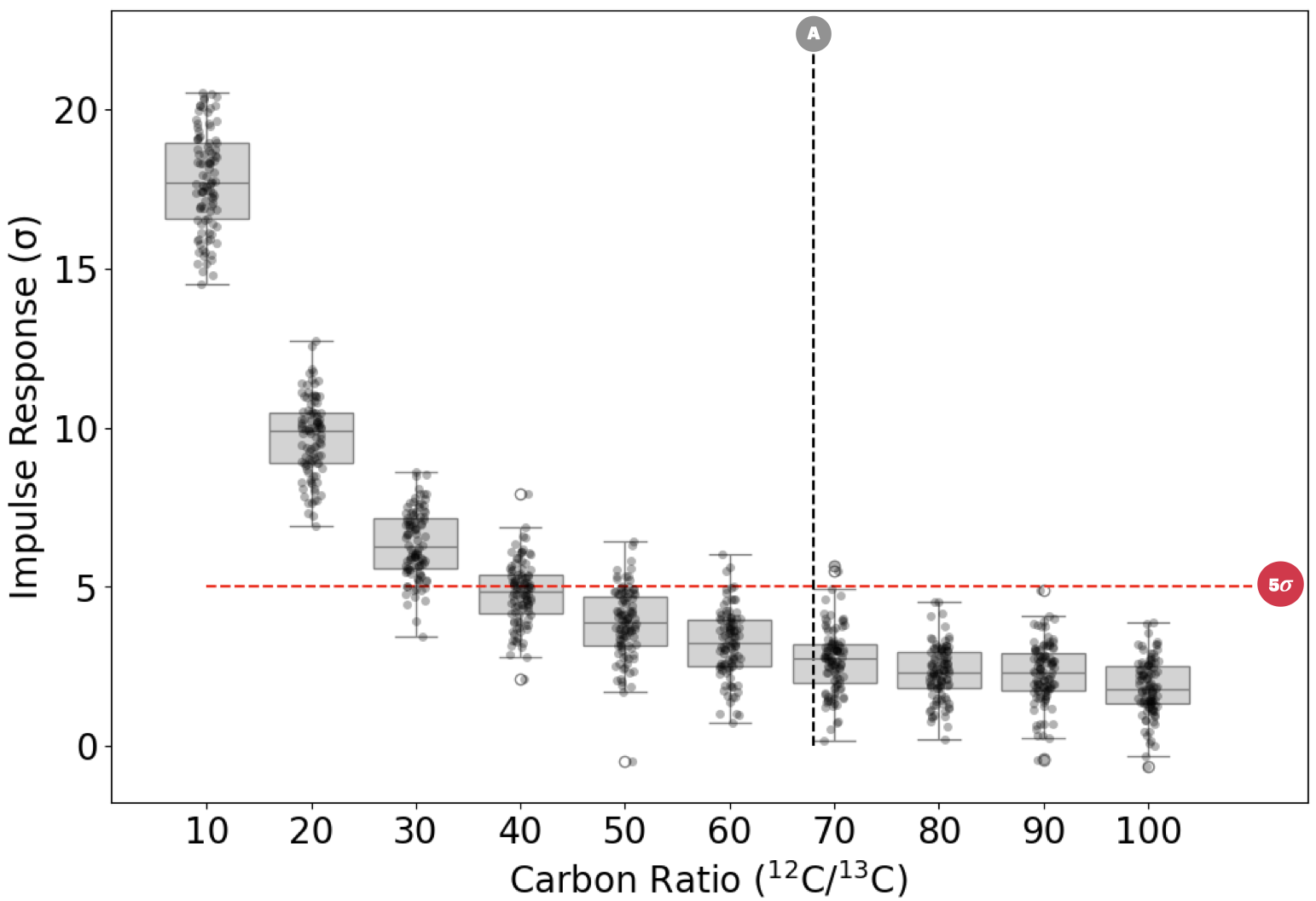}
    \caption{Summed isotopologue stack and filter impulse response distributions for the aggregate \ce{^13C} isotopologues of cyanonaphthalene (n=100). Each stack simulation includes a total of 11 unique isotopologues across both isomers. Carbon isotope ratios are based off of observed abundances for the parent species. The vertical dashed line (A) represents the local bulk \ce{^12C}/\ce{^13C} ratio of $\sim$69, and the dashed red line is the adopted 5$\sigma$ impulse response threshold of detection.}
    \label{fig:cnns}
\end{figure}

\subsubsection{Cyanonaphthalenes} There are 11 unique singly-substituted \ce{^13C} isotopomers of 1- and 2-cyanonaphthalene. Fig.~\ref{fig:cnns} shows the results for aggregate signal retrieval in our simulations. We find that at \ce{^12C}/\ce{^13C} = 35, we expect an average impulse response of 5$\sigma$, although this is clearly dependent on the exact noise properties of the spectra, with a non-trivial number of runs giving responses between 3--5$\sigma$.  At \ce{^12C}/\ce{^13C} $\lessapprox$30, a robust $>$5$\sigma$ detection should be expected regardless. Given the local bulk \ce{^12C}/\ce{^13C} ratio of $\sim$\,69 in TMC-1 \citep{colzi_carbon_2020}, a non-detection in the data would clearly not be consistent with an enhanced abundance of \ce{^13C} from a top-down, circumstellar inheritance scenario, which would be a ratio below that of local bulk \ce{^12C}/\ce{^13C}.  As noted above, this would not entirely rule out a top-down inheritance, as an enhanced \ce{^13C} ratio is not mandatory in this scenario, but it would be suggestive.  On the other hand, a detection at this level would strongly imply an inheritance scenario.

\begin{figure}
    \centering
    \includegraphics[width=1\columnwidth]{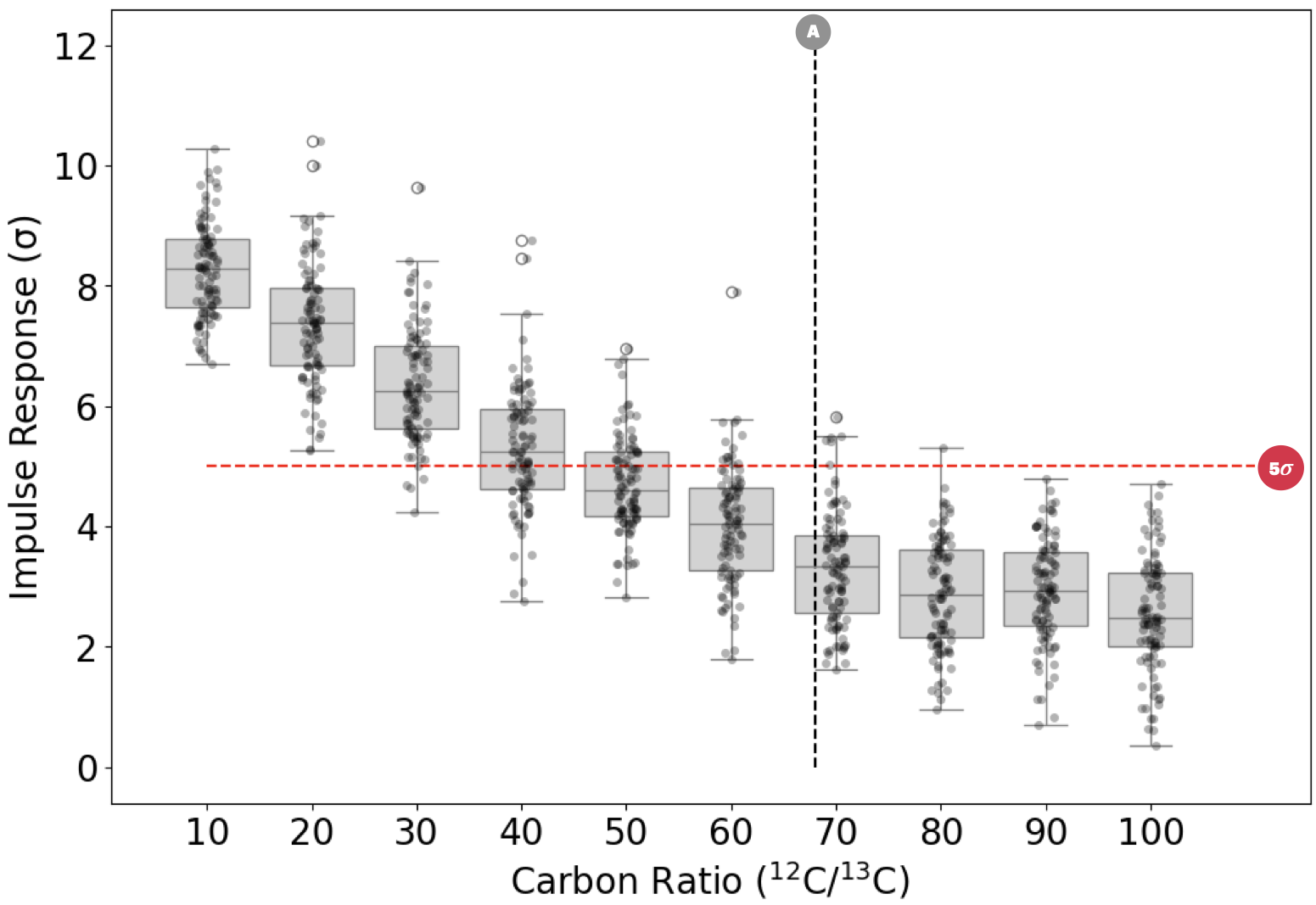}
    \caption{Summed isotopologue stack and filter impulse response distributions for the aggregate \ce{^13C} isotopologues of 1, 2-, and 4-cyanopyrene. Each stack simulation includes a total of 45 unique isotopologues across all four isomers (17 isotopomers each for 1- and 4-cyanopyrene, and 11 for 2-cyanopyrene), with 100 iterations per ratio. Carbon isotope ratios are based off of observed abundances for the unsubstituted parent species. The vertical dashed line (A) represents the local bulk \ce{^12C}/\ce{^13C} ratio of $\sim$69, and the dashed red line is the adopted 5$\sigma$ impulse response threshold of detection.}
    \label{fig:cpys}
\end{figure}

\subsubsection{Cyanopyrenes} 
There are 45 unique isotopomers of 1-, 2-, and 4-cyanopyrene.  Fig.~\ref{fig:cpys} shows the results for aggregate signal retrieval in our simulations.  The overall trend is very similar to the cyanonaphthalenes, which is not unexpected -- the total abundances of cyanonaphthalene ($1.44\times10^{12}$\,cm$^{-2}$; \citealt{mcguire_detection_2021}) and cyanopyrene ($3.7\times10^{12}$\,cm$^{-2}$; \citealt{wenzel_discovery_2025,wenzel_detection_2024}) in TMC-1 are extremely similar.  The spread in responses, however, is much larger for cyanopyrene retrieval, which we speculate is because the isotopologues of cyanopyrene posses many more, much weaker transitions than those of cyanonaphthalene, increasing the impact of noise. Still, we again find that at \ce{^12C}/\ce{^13C}\,$\lessapprox$\,30, a robust $>$\,5$\sigma$ detection should be expected regardless.

\section{Discussion}
Beyond the \ce{^12C}/\ce{^13C} ratios, there is evidence suggesting likely bottom-up, \emph{in situ} formation of the small PAHs being found in TMC-1. A number of modeling studies have investigated the photodestruction of PAHs in the ISM, particularly their stability in regions where strong IR emission has been attributed to PAH vibrational modes. For example, \citet{allain_photodestruction_1996} developed a model to compare the rate of acetylene loss to the accretion rate of carbon to PAHs, showing that species with more than 50 carbon atoms may be able to survive the radiation in different UV irradiated regions, but smaller PAHs like coronene or ovalene are rapidly destroyed. Later work by \citet{montillaud_evolution_2013} indicated this limit may be even lower, with PAHs of $\sim$30 carbons or more surviving. If correct, then the detections of cyanonaphthalene (10 carbon atoms in the rings), cyanopyrene (16 carbon atoms in the rings), and most recently cyanocoronene (24 carbon atoms in the rings) would suggest that circumstellar formation is not likely to be responsible for their presence, as their survival from a CSE to a cold molecular cloud would be highly unlikely. Isotopic analysis of PAHs, specifically naphthalene and pyrene, in samples returned from asteroid Ryugu by \citet{zeichner_polycyclic_2023} also favor a cold interstellar formation scenario, showing substitution patterns consistent with formation at $\sim$10\,K. That said, recent experimental and theoretical work studying the photodestruction of naphthalene has revealed potential stabilization pathways through recurrent fluorescence that may enhance their survivability between generation in a CSE and incorporation into a dark cloud \citep{stockett_efficient_2023}.  A direct measurement of the \ce{^12C}/\ce{^13C} ratio of these species in TMC-1 using the techniques outlined here, or at the very least a constraining limit, would therefore go a long way in clearing up the mystery.

Indeed, \ce{^13C} fractionation patterns have long been used to infer provenance and chemical formation pathways. \citet{burkhardt_detection_2018} reported the first interstellar detections of six \ce{^13C} bearing isotopologues of \ce{HC7N} and highlighted the utility of isotopic composition for studying the low temperature formation pathways that may be responsible for certain carbon-chain species in TMC-1. From direct observations of \ce{HC5N} and \ce{HC7N} isotopologues in TMC-1, chemical signatures of three distinct formation mechanisms were delineated using the relative isotopic ratios of all \ce{^13C} isotopomers of the parent. Based on isotopomer abundance ratios for both \ce{^13C} and \ce{^15N}, the lack of isotopomer abundance variations at the nitrile carbon may indicate that direct formation through the addition of \ce{CN} to a longer existing carbon chain is unlikely to be an important route in their formation. \textcolor{black}{For other molecular species, a} variation in isotopomer abundance ratios may be indicative of differing formation routes, despite being in the same molecular family. 

Chemical modeling studies have been used in this vein to elucidate reaction mechanisms and probe the early chemistry in PAH formation. Using a gas-grain chemical network model, \citet{furuya_carbon_2011} suggested that the carbon isotope ratios for gas-phase species are correlated based on whether formation begins from a carbon atom or from a CO molecule; the \ce{^12C}/\ce{^13C} ratios for species formed from a carbon atom are larger than the carbon elemental abundance ratio, while those formed from the CO molecule are smaller \citep{furuya_carbon_2011}. This dependency is closely linked to exothermic isotopic exchange reactions that may be responsible for isotopic fractionation patterns specific to low temperature environments.  Indeed, at the low temperatures of TMC-1, it is possible to see enrichment of \ce{^13C} in some species as well, potentially confusing the interpretation of inheritance from CSE. The key reaction from \citet{furuya_carbon_2011} is:
\begin{equation}
    \ce{^13C+ + ^12CO <=> ^12C+ + ^13CO + 35 K}.
\end{equation}
Under interstellar conditions at temperatures substantially higher than 35\,K, the small exothermicity of the forward pathway will have a negligible effect on the overall fractionation between CO and \ce{C+}.  At 10\,K, however, the backward reaction is highly inefficient, resulting in an enhanced \ce{^12C}/\ce{^13C} ratio in \ce{C+} and a lower ratio in CO.  Species formed primarily through reactions beginning with \ce{C+}, therefore, may show an even \textcolor{black}{higher} \ce{^12C}/\ce{^13C} ratio than expected from the local, bulk gas. The dominant pathways to PAH formation in current models are through reactions which originate with atomic carbon, either as C or \ce{C+} \citep{byrne_astrochemical_2023}, thus making any detection of a \textcolor{black}{depleted} \ce{^12C}/\ce{^13C} ratio in PAHs in TMC-1 even more suggestive of a top-down inheritance scenario.

\section{Conclusions}
We have shown that the application of spectral line stacking and matched filtering can successfully be implemented in the accurate recovery of aggregate PAH isotopologue signals. The final signal that is recovered cannot be artificially inflated by a high density of buried molecular lines, nor by stacking simple white Gaussian noise. With a better understanding of the limits of this technique, \textcolor{black}{we are also able} to constrain the pool of possible astronomical targets and prioritize those with strong enough aggregate signal  in future laboratory studies. Additionally, using \ce{HC9N} as a case study, we have also shown that preliminary constraints can be placed on the carbon isotope ratio based on signal recovery curves. The lower limit of the carbon isotope ratio for \ce{HC9N} estimated from our presented analysis is consistent with previous estimates of cyanopolyynes in TMC-1 that are thought to be formed \textit{in situ} in the cloud itself. \textcolor{black}{Based on the work presented here, we suggest that if there is a significant level of \ce{^13C} enrichment in cyanonaphthalene or cyanopyrene in TMC-1, their molecular signal can be recovered using the methods presented here. Future work will discuss this technique using the laboratory spectra of PAH isotopologues.}

\section*{Acknowledgments}
B.A.M. and C.X. gratefully acknowledge support of National Science Foundation grant AST-2205126. B.A.M and M.D. gratefully acknowledge support from Schmidt Family Futures. G.W. and B.A.M. acknowledge the support of the Arnold and Mabel Beckman Foundation Beckman Young Investigator Award and Schmidt Family Futures. I.R.C. acknowledges support from the Natural Sciences and Engineering Research Council of Canada (grant RGPIN-2022-04684), the Canada Foundation for Innovation and the B.C. Knowledge Development Fund. The National Radio Astronomy Observatory is a facility of the National Science Foundation operated under cooperative agreement by Associated Universities, Inc. This paper makes use of the following GBT data: AGBT17A-164, AGBT17A-434, AGBT18A-333, AGBT18B-007, AGBT19B-047, AGBT20A-516, AGBT21A-414, 21B-210, and AGBT24A-124. 

\bibliography{refs.bib}
\bibliographystyle{aasjournal}

\clearpage

\appendix
\section{An Empirical Estimate of the GOTHAM Line Density}
\label{appendix:linedensity}

As shown in Section \ref{sec:results}, the impact of spectral confusion on stacking and matched filtering depends on the density of weak, potentially unidentified features in the data. We have therefore attempted to empirically characterize the occurrence rate of spectral lines in the GOTHAM survey. Previous discussions of the line density below the nominal noise level in GOTHAM were based on simple extrapolations from the observed density of bright lines \citep[e.g.][]{mcguire_detection_2021}. Here, we use the public GOTHAM data release \citep{xue_molecular_2025} to derive an empirical, completeness-limited estimate of the cumulative line density as a function of intensity threshold.

We loosely follow the methodology of \citet{mcguire_detection_2021} with a few important distinctions. We define the cumulative line density at an intensity cutoff $I$ in terms of the fraction of the surveyed bandwidth occupied by all detectable features with peak intensity greater than or equal to $I$. For a line with an assumed FWHM linewidth $\Delta v_{\rm FWHM}$ at sky frequency $\nu_i$, the corresponding frequency width is
\begin{equation}
\Delta \nu_i = \nu_i \frac{\Delta v_{\rm FWHM}}{c}.
\end{equation}
The cumulative line density is then
\begin{equation}
\eta(I) = \frac{\sum_i \Delta \nu_i}{B_{\rm eff}(I)},
\end{equation}
where the sum is taken over all peaks satisfying the conditions \textcolor{black}{that the peak brightness is greater than intensity cutoff $I$} 
\begin{equation}
T_{{\rm mb},i} \ge I
\end{equation}
and \textcolor{black}{the intensity cutoff must be above a minimal multiple $n_{\sigma}$ of the local noise $\sigma_{{\rm local}}$}
\label{eqA4}
\begin{equation}
I \ge n_\sigma \sigma_{{\rm local},i},
\end{equation}
and where $B_{\rm eff}(I)$ is the effective searchable bandwidth for that same cutoff. In other words, both the numerator and denominator are restricted to the portion of the survey for which a line of intensity $I$ would be observable under the adopted detection criterion. This matching is important and differs from the \citet{mcguire_detection_2021} analysis. If only a fraction of the survey reaches the sensitivity required to detect a line of intensity $I$, then only that same fraction of the bandwidth should contribute to the normalization.

To estimate $\eta(I)$, we searched the GOTHAM spectrum for positive local maxima \textcolor{black}{above a signal-to-noise threshold of 5~$\times~\sigma_{{\rm local}}$} and imposed a minimum separation between neighboring peaks of 0.5 times the assumed FWHM of $\sim$0.4~km~s$^{-1}$. We then evaluated the cumulative density as a function of intensity cutoff. \textcolor{black}{The minimum cutoff intensity is set by Equation \ref{eqA4}4 such that some minimal part of the GOTHAM bandpass reaches the required sensitivity for the measured line density value to stabilize.} At high cutoff intensities the curve then steepens because only a small number of bright lines remain, and the cumulative distribution becomes dominated by small-number statistics. The approximately power-law segment between these two limits is therefore the most useful regime for characterizing the line density behavior of the survey. As noted in \citet{mcguire_detection_2021}, there is not a clear physical motivation for this power-law behavior, but it appears to fit the data well and thus is empirically motivated.

\begin{figure}
    \centering
    \includegraphics[width=0.7\linewidth]{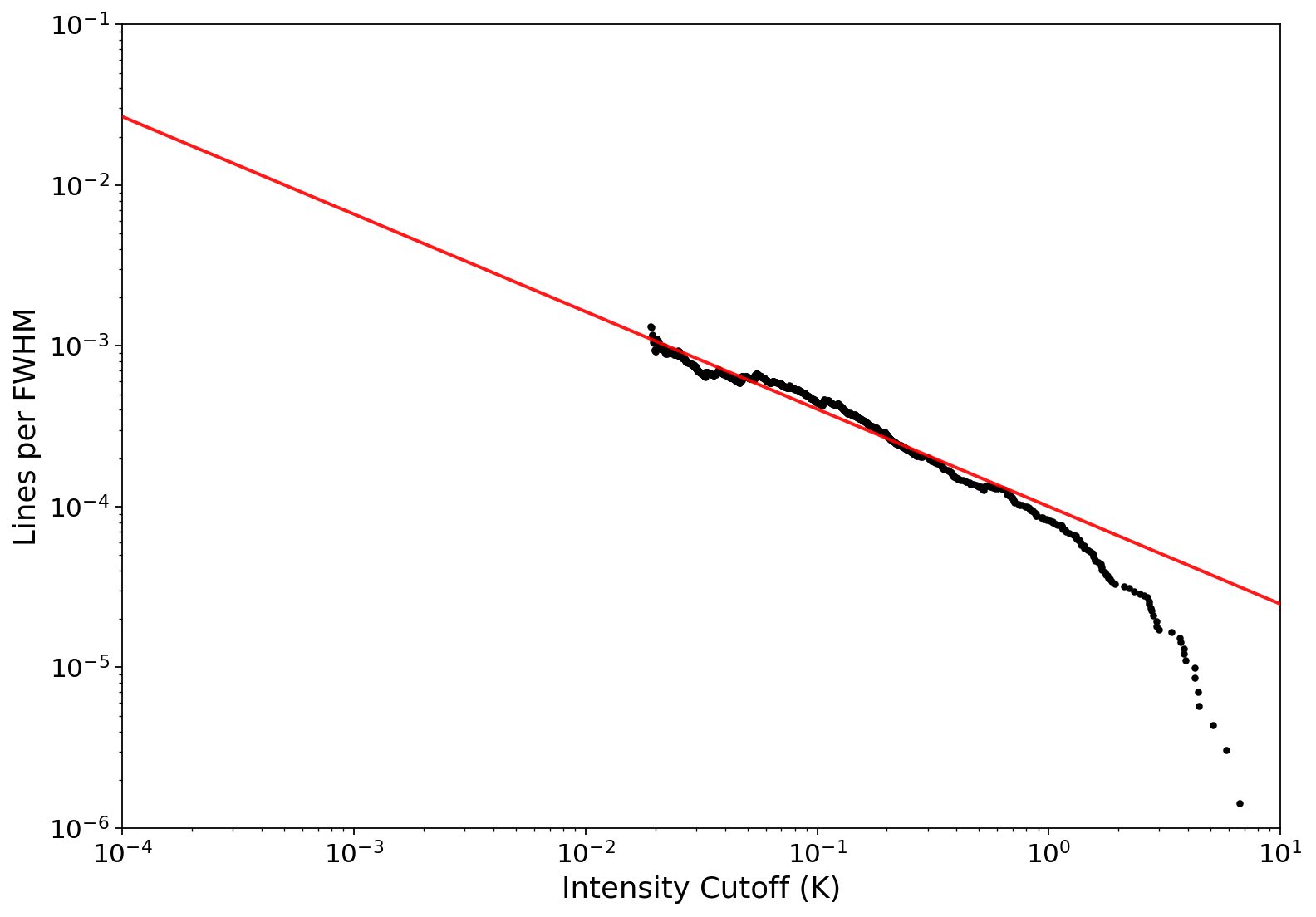}
    \caption{Empirical cumulative line density of the GOTHAM spectrum as a function of intensity cutoff. A power-law fit is overlaid, where the fit is restricted to the completeness-limited, approximately power-law portion of the curve.}
    \label{fig:linedensity_all}
\end{figure}

Figure~\ref{fig:linedensity_all} shows the resulting cumulative line-density curve. The low-intensity flattening seen in Figure S9 of \citet{mcguire_detection_2021} is suppressed once the normalization is restricted to the effective bandwidth accessible at each intensity threshold, suggesting that the apparent flattening was primarily a completeness effect rather than a true change in the underlying line occurrence rate. We therefore caution against interpreting a single extrapolated survey-wide number below the noise level as a physically meaningful ``line density of TMC-1.'' 

\begin{figure}
    \centering
    \includegraphics[width=0.7\linewidth]{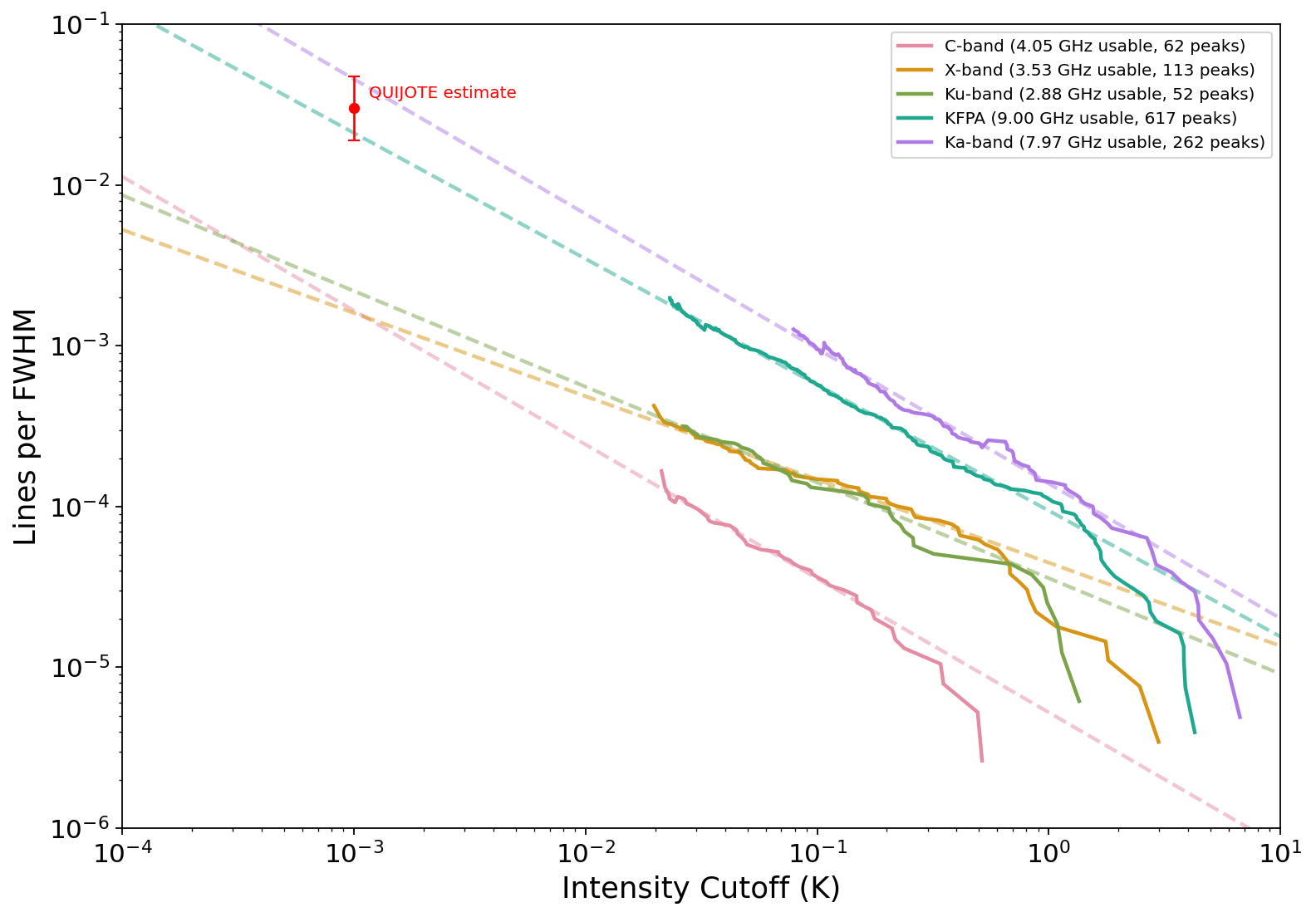}
    \caption{Same as Figure~\ref{fig:linedensity_all}, but computed separately for each major GOTHAM receiver band. The cumulative line density increases toward higher frequency, demonstrating that the weak-line population in GOTHAM is not well described by a single survey-wide extrapolation. Dashed lines show power-law fits to the approximately linear portion of each curve in log-log space.}
    \label{fig:linedensity_bands}
\end{figure}

The true density of weak features is likely both sensitivity-limited and frequency dependent, as there is a correlation between molecular size, partition function, and frequency location of brightest lines for a given excitation temperature. To examine that frequency dependence explicitly, we repeated the same analysis separately for the GBT receiver bands used by GOTHAM. For the purposes of simple analysis, we divided them as follows: C ($<8$ GHz), X ($8$--$12$ GHz), Ku ($12$--$16$ GHz), KFPA ($17.5$--$27$ GHz), and Ka ($>27$ GHz). The results are shown in Figure~\ref{fig:linedensity_bands}. The line density increases systematically toward higher observing frequency, with the Ka-band spectrum exhibiting the highest density of weak features and the C-band spectrum the lowest. This behavior is qualitatively expected, and we conclude that there is no single line-density curve that is representative of TMC-1 at all frequencies.

This band dependence is important for proper comparison to the higher frequency QUIJOTE survey of TMC-1 \citep{cernicharo_quijote_2022}. QUIJOTE only has moderate overlap with the GOTHAM frequency range, and limited portions of the survey are only publicly available in figure form, making a direct line-by-line comparison difficult. We attempt a rough comparison via visual examination of the spectra from \citet{cernicharo_discovery_2021}. There, the line density appears to be roughly one line per 1~MHz at $\sim$35~GHz. The visual linewidth appears to be closer to $\sim$0.6~km~s$^{-1}$, but this is roughly consistent with the GOTHAM observed linewidth of $\sim$0.4~km~s$^{-1}$ when considering the much wider channel width of QUIJOTE at 38~kHz. Thus we estimate that QUIJOTE has roughly a line density of one line per 30~FWHM at a 5$\sigma$ cutoff of 1~mK, and we assign 0.2~dex of uncertainty. Overplotting this on our analysis, we find it to be in reasonable agreement with extrapolation from Ka band GOTHAM observation \textcolor{black}{(Fig.\ref{fig:linedensity_bands})}.

We therefore conclude that GOTHAM remains line sparse in the regime of directly detected lines, but the cumulative occurrence rate of weaker features rises steadily below the nominal noise level, and that rise is strongly frequency dependent. This justifies treating the hidden-line population in our synthetic stress tests as uncertain and suggests not to over-interpret any apparent numerical correspondence between the measured GOTHAM line density and the turnover point seen in Figure~2.

\section{False Positive/Negative Impulse Response Distributions}
\begin{figure*}[htb!]
    \centering
    \includegraphics[width=1\linewidth]{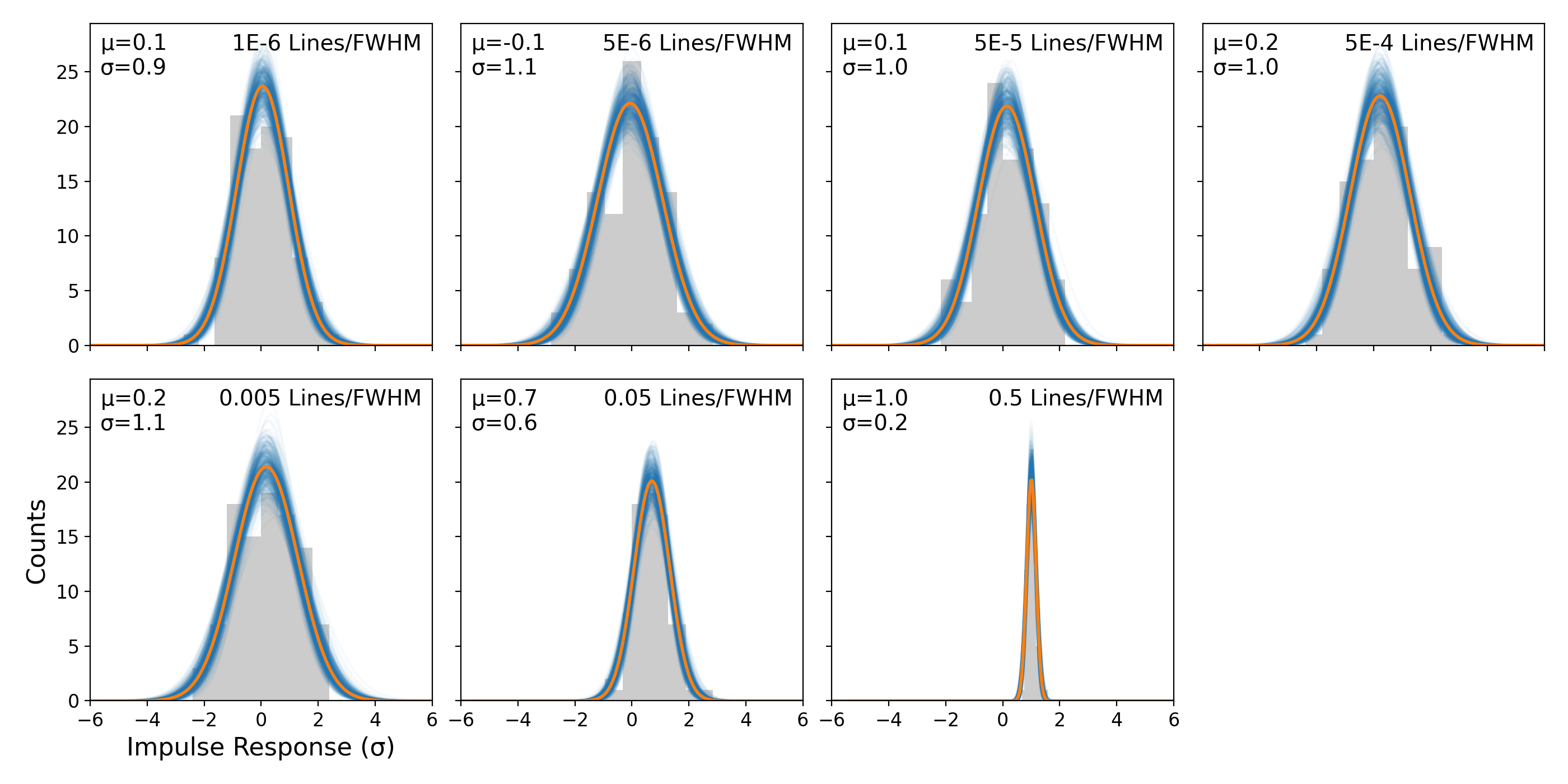}
    \caption{False positive gaussian distributions of 1-cyanonaphthalene impulse responses from noise-only synthetic spectra (n=100) at varying line densities.}
    \label{fig:gauss_pos}
\end{figure*}
\end{document}